\documentclass[a4paper,fleqn]{cas-dc}
\usepackage{siunitx}
\usepackage{commath}
\usepackage{amsmath}
\usepackage{gensymb}

\usepackage[numbers,sort&compress]{natbib}
\def\tsc#1{\csdef{#1}{\textsc{\lowercase{#1}}\xspace}}
\tsc{WGM}
\tsc{QE}
\tsc{EP}
\tsc{PMS}
\tsc{BEC}
\tsc{DE}

\begin{document}
\let\WriteBookmarks\relax
\def\floatpagepagefraction{1}
\def\textpagefraction{.001}
\shorttitle{Intermediate energy proton irradiation and testing facility}
\shortauthors{S Jepeal}

\title [mode = title]{An accelerator facility for intermediate energy
  proton irradiation and testing of nuclear materials}

\author[1]{Jepeal S. J.}
\author[1]{Danagoulian, A.}
\author[1]{Kesler, L. A.}
\author[1]{Korsun, D. A.}
\author[1]{Lee, H. Y.}
\author[2]{Schwartz, N.}
\author[1]{Sorbom, B. N.}
\author[1]{Velez Lopez, E.}
\author[1]{Hartwig. Z. S.}

\address[1]{Department of Nuclear Science and Engineering, Massachusetts Institute of Technology, Cambridge MA USA}
\address[2]{Department of Mechanical Engineering, Massachusetts Institute of Technology, Cambridge MA USA}

\begin{abstract}
The bulk irradiation of materials with 10 -- 30 MeV protons promises to advance the study of radiation damage for fission and fusion power plants. Intermediate energy proton beams can now be dedicated to materials irradiation within university-scale laboratories. This paper describes the first such facility, with an Ionetix ION-12SC cyclotron producing 12~MeV proton beams.  Samples are mm-scale tensile specimens with thicknesses up to \SI{300}{\micro\m}, mounted to a cooled beam target with control over temperature. A specialized tensile tester for radioactive specimens at high temperature ($500+$\degree C) and/or vacuum represents the conditions in fission and fusion systems, while a digital image correlation system remotely measures strain. Overall, the facility provides university-scale irradiation and testing capability with intermediate energy protons to complement traditional in-core fission reactor and micro-scale ion irradiation. This facility demonstrates that bulk proton irradiation is a scalable and effective approach for nuclear materials research, down-selection, and qualification.
\end{abstract}

\begin{highlights}
\item A first-of-its-kind bulk materials irradiation facility has been built around a superconducting cyclotron producing 12~MeV protons.
\item Miniature samples of 100-\SI{300}{\micro\meter} can be irradiated uniformly at dose rates of 0.1+ dpa/day with $\pm$10\degree C temperature control.
\item A high temperature tensile tester with a custom furnace and environmental chamber allows mechanical testing of fusion and fission materials in the relevant operating conditions.
\item An optical strain measuring system based on digital image correlation allows non-contact measurements of strain in miniature tensile samples.
\item Mechanical test of miniature samples are shown to allow highly reproducible measurements of strength ($<$ 2\% variation) and ductility ($\sim$ 3\% variation).
\end{highlights}


\maketitle

\section{Introduction}
\label{sec:introduction}

Realizing the potential of future fusion and nuclear fission
power plants will require the development, down selection, and
qualification of materials that satisfy the demanding design and
operational requirements in the extreme environments
of these systems. For example, the materials surrounding the core of a
fusion power plant are expected to receive total neutron fluxes of
10$^{14}$ to 10$^{15}$ cm$^{-2}$~s$^{-1}$, equivalent to roughly 15
displacements per atom (DPA) per operation year. Advanced Gen IV fission reactors are exploring operating scenarios with in-core materials reaching lifetime doses in the hundreds of DPA. Such high levels of
neutron exposure can cause undesirable property changes such as
hardening, embrittlement, and decreases in thermal conductivity,
which must be mitigated if these technologies are to be viable
solutions to global climate change and growing demand for energy.
Advancing materials such that they can survive the hostile radiation
environments of fusion and advance fission requires new experimental
techniques that provide high fidelity irradiated material responses on
accelerated timelines. We have proposed such a technique -
intermediate energy proton irradiation, or IEPI - that uses 10~MeV to
30~MeV proton beams to rapidly irradiate bulk material specimens that
are suitable for macroscopic testing to directly extract engineering
properties \cite{jepeal2020}.

The objective of this paper is to introduce the first university facility - The
Vault Laboratory at MIT - that has the dedicated capability to conduct
IEPI and provide technical insights into the equipment,
infrastructure, and processes used to execute the
technique. Experimental validation results, both for the irradiation
and post-irradiation activities, are presented to demonstrate confidence
in the underlying processes and equipment. Based on this facility and the
initial IEPI experience, recommendations for optimized future facilities
can be extracted.

This paper is structured as follows: the remainder of
Section~\ref{sec:introduction} provides an overview of the IEPI
technique and the enabling technology: compact superconducting
cyclotrons; Section~\ref{sec:vault} provides an overview of the Vault
Laboratory with a focus on the layout and infrastructure required for
IEPI; Section~\ref{sec:irradiation} presents details of the present
hardware and capabilities for IEPI in the Vault Laboratory; Section~\ref{sec:pie} presents details of the mechanical evaluation equipment for IEPI samples.
Section~\ref{sec:conclusion} concludes with an outlook for the
IEPI technique and future facilities.

\subsection{Intermediate energy proton irradiation}
\label{subsection:iepi}

Two techniques are predominantly employed at present to study material
evolution under high doses of radiation in future fission and fusion power plants: neutron irradiation in test
reactors and micro-scale ion bombardment in accelerators. These
techniques have substantially advanced our understanding of material
performance, evolution, and lifetime under irradiation. However, test reactors irradiation is limited by high costs and long experimental times, and micro-scale ion irradiation cannot directly provide engineering property measurements.  It has been recently recognized that
a third irradiation technique, which we call intermediate energy
proton irradiation or IEPI, has the capability to meet emerging needs of nuclear materials development.
\cite{rayaprolu2016,zinkle2018, jepeal2020}. IEPI enables high fidelity measurement of radiation damage in macroscopic samples, and is expected to complement the existing in-core reactor and ion irradiation techniques that presently form the vast majority of radiation damage studies. A comprehensive
scientific and engineering analysis of the IEPI technique can be found
in \cite{jepeal2020}.

\begin{figure}
  \centering
  \includegraphics[width=0.48\textwidth]{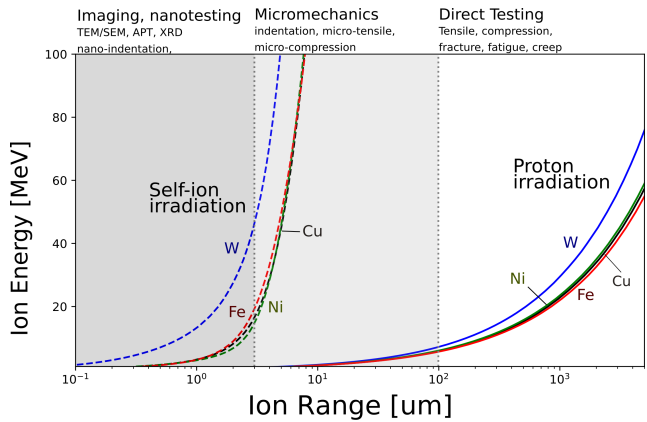}
  \caption{The range of protons and self-ions (an example of heavy ions) in four common elemental metals are shown. Protons above 10~MeV produce bulk-scale (>\SI{100}{\micro\meter}) damage, enabling direct testing methods that are infeasible for self-ion irradiation. The ranges of all ions were computed with SRIM \cite{srim}.}
  \label{fig:ranges}
\end{figure}

At the core of the IEPI technique is the use of 10~MeV to 30~MeV
(``intermediate energy'') proton beams that have ranges in materials
roughly two orders of magnitude higher than either few-MeV protons or
self-ions. Figure~\ref{fig:ranges} shows a representative set of ion ranges in materials for some common ion species and materials as a function of beam energy, as calculated with SRIM~\cite{srim}. Intermediate
energy proton have ranges in these materials on the order of 0.1--1.0~mm,
enabling irradiation in macroscopic material specimens that can be
directly tested for engineering properties. Low-energy proton or 
heavy ion (including self-ion) irradiation of energies up to $\sim$100 MeV achieve irradiation depths only on
the order of 10~nm to \SI{10}{\micro\meter}, requiring challenging
post-irradiation experimental techniques and complicating the
extraction of macroscopic engineering data.  While sources of heavy ions above 100 MeV exist, damage by these swift heavy ions is strongly influenced by the rapid ionization of the irradiated material, making them a poor tool for simulating neutron damage, and limiting their use primarily to simulating damage by high energy fission products within nuclear fuel~\cite{zinkle2018, Matzke2000}.

\subsection{Opportunities with superconducting cyclotrons}
\label{subsec:opportunities}
While a variety of established accelerator research facilities have access to intermediate energy protons~\cite{Voyles2019,Gupta2020}, there are complication regarding IEPI that have prevented its widespread use.  Damage rates for IEPI can be 1--2 orders of magnitude lower than heavy ion irradiation~\cite{jepeal2020}, requiring days of irradiation to reach dpa-levels of damage. Such long irradiation times prevent extensive use of IEPI in shared accelerator user facilities with high hourly fees and limited availability. Extremely high heat fluxes created by the proton beam heating~\cite{jepeal2020} challenge the ability to control the irradiation temperature, requiring highly customized cooling and temperature monitoring. Nuclear reactions caused by the proton beam require a laboratory to be equipped to shield personnel from radiation during experiments and require handling low-level radioactive material after irradiation~\cite{jepeal2020}. Because of this need for long irradiation times and specialized capabilities, IEPI is not well suited for existing shared accelerator facilities that are built to tailor to diverse applications, but instead necessitates a dedicated experimental facility. 

The demand for sources of medical radioisotopes other than
nuclear reactors has stimulated significant investment and innovation
in compact ($\sim$1~m$^{3}$), moderate-to-high current (\SI{10}{\micro\ampere}
to 10~mA), intermediate energy (10 MeV to 30 MeV)
\cite{smirnov2016} superconducting cyclotrons. Advances in high field superconducting magnets,
liquid-free conduction cooled cryogenic technology, and accelerator
beam dynamics has led to the demonstrations and deployments of the
first real-world systems produced by commercial companies
\cite{vincent2016}. Design studies have been conducted - with
prototypes now being fabricated - for machines that would extend the
beam energy up to 70~MeV with external proton beam currents in the
mA range.  \cite{smirnov2019,kelly2019}. As discussed in detail
by Jepeal \textit{et al.}, such machines would be ideal for maximizing
the capabilities of the IEPI technique \cite{jepeal2020}.

The beam characteristics of medical isotope cyclotrons are well-suited
for the irradiation of materials. The small footprint and weight of
the cyclotron and support equipment - coupled to the low power
consumption due to the main confining superconducting magnet - enables
relatively straightforward installation in university-scale research
laboratories. Deployments in university research labs - as opposed to
medical facilities - allow the machines to be dedicated full-time to
materials research and provide significantly more operational
flexibility. The absence of liquid helium coolant eliminates major
infrastructure and operating expenses while increasing personnel
safety. The commercial provenance of the machines results in high beam
availability, straightforward maintenance, and robust control systems
which are often lacking in research-grade accelerators and hinder the
use of ion accelerators for materials research.

\section{The Vault Laboratory at MIT}
\label{sec:vault}

This section provides an overview of the Vault Laboratory at MIT,
which is co-operated by the Department of Nuclear Science and
Engineering and the Plasma Science and Fusion Center. The laboratory
specializes in the use of accelerators to advance the science of
materials for fission and fusion systems principally through the
modification and measurement of materials with accelerated ion beams.

The Vault Laboratory contains three accelerators: First, a Newton
Scientific Instruments 2~MV tandem ion accelerator (DANTE) capable of
producing proton and deuteron beams. The machine has been used in the
development and qualification of novel diagnostic techniques for
materials inside of magnet fusion devices \cite{hartwig2013,
  hartwig2014_jnm, kesler2017}, investigation into plasma-surface
interactions with fusion materials \cite{wright2011}, and is presently
being used for irradiation of high temperature superconductor intended
for use in high-field magnets for accelerators and fusion devices
\cite{sorbom2017}. Second, a deuterium-tritium
neutron generator is capable of producing 14.1~MeV neutrons at rates up
to 3$\times$10$^{8}$ neutrons per second. The generator is used primarily for radiation shielding studies, neutron detector
development, and validation of new Monte Carlo particle transport
methods \cite{hartwig2014_nima} and experimental data acquisition
frameworks \cite{hartwig2016}. The third accelerator is an Ionetix
ION-12SC proton superconducting cyclotron. This machine is
a key component of the IEPI facility paper and is described in detail in
Section~\ref{subsec:cyclotron}. In addition to materials irradiation, this accelerator has also been utilized in a novel radiographic system measuring the atomic number and density of sample materials with applications in cargo scanning and nuclear security\cite{lee2019multiple}.

The following sections provide a brief overview and description of the
experimental space and infrastructure required to efficiently and
safely operate the cyclotron for intermediate energy proton
irradiation of materials.

\subsection{Experimental space layout}
\label{subsec:layout}

\begin{figure}
  \centering
  \includegraphics[width=0.48\textwidth]{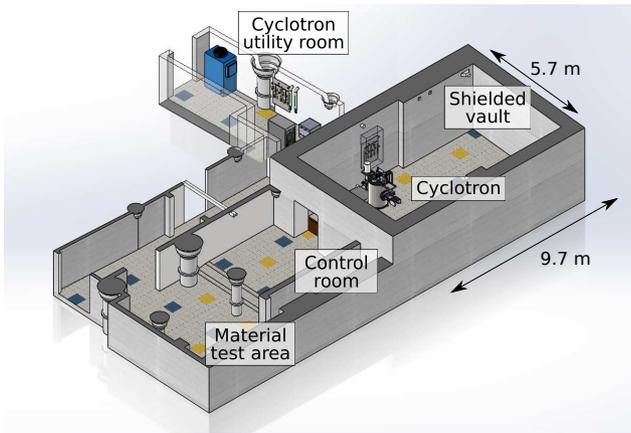}
  \caption{CAD rendering of the Vault Laboratory layout, showing only
    the accelerators and equipment relevant to this paper: the
    cyclotron within the meter-thick shielded vault; the cyclotron
    support systems; and the control room and materials test area.}
  \label{fig:layout}
\end{figure}

A CAD rendering of the Vault Laboratory is shown in
Figure~\ref{fig:layout} with only the superconducting cyclotron and
associated equipment relevant to this paper shown and described for
simplicity. The space is divided primarily into three main areas: a
shielded vault; accelerator bays; and a control room.

The shielded radiation vault accommodates safe production of high doses
of prompt and delayed radiation from accelerator operations. Because
the cyclotron contains an internal beam target, it is located within
the vault for radiation shielding effectiveness.

The accelerator bays are adjacent to the vault with penetrations
through the shield wall. One of these accelerator bays (the cyclotron
utility room) hosts the equipment required to operate the cyclotron: a
rack containing the control system, RF power supplies, and magnet
power supplies; a helium cryocompressor for the cryopump and coldhead
system that maintains high vacuum within the cyclotron and keeps the
cyclotron magnet cold through conduction cooling; the large water
chiller, which provides a secondary pure distilled water loop to cool
the cyclotron target, RF system, and cryocompressor; the hydrogen and
nitrogen gas bottles; and a water and gas distribution
manifolds. Penetrations through the vault shield wall enable access
into the vault for the electrical, RF, cooling water, and gases; they
also enable prompt radiation from cyclotron operations to be
collimated and measured for precise neutron and gamma spectroscopy if
desired.

The large room next to the vault serves as the remote control room for
all of the accelerators in the laboratory, as well an area dedicated
to material specimen preparation and post-irradiation experiments. The
post-irradiation experimental equipment described in
Sections~\ref{sec:pie} is located here.

\subsection{The shielded radiation vault}
\label{subsec:vault}

The core of the laboratory is the large (9.7~m $\times$ 5.7~m $\times$
3.5~m high) shielded vault. The vault is surrounded by 0.82~m thick
high density rebarred concrete shield walls and ceiling with
penetrations for accelerator beam access into and out of the
vault. Multiple right right-angle penetrations through the shield wall
provide access for instrumentation wires, control signals, and other
utilities without radiation leakage. A wide 0.64~m thick steel and
high density concrete shield door provides physically access to the
vault. The door is suspended on overhead rails and powered by a large
motor for movement and slots into a 0.1~m deep trench to eliminate
radiation leakage.

An important aspect of siting the cyclotron within the Vault
Laboratory was assessing the prompt radiation shielding capabilities
of the vault to ensure radiation safety for personnel during
operations. There are three unique challenges for intermediate energy
proton irradiation of materials that require careful
evaluation. First, the nominal cyclotron beam energy of 12~MeV exceeds
the nuclear reaction thresholds for almost all elements, leading to
substantially higher prompt gamma and neutron radiation compared with
traditional proton accelerators operating at a few MeV. Of particular
concern are the (p,n) reactions that lead to high energy neutron
production. Second, the focus of bulk irradiation of materials requires producing thick-target yields of gammas and neutrons from the material specimen. Third, a wide range of materials for fission
and fusion systems are expected to be irradiated in the facility,
 such as beryllium and tungsten, requiring careful
assessment of the prompt radiation produced from each target
material.

To evaluate the radiation shielding, a detailed three dimensional model of the Vault Laboratory was constructed in MCNP (Version 5-1.60), a leading Monte Carlo particle transport code \cite{mcnp}. The model includes all major geometry features, detailed material compositions, and several particle sources to correspond to different irradiation experiments. Importantly, the particle source can be customized to directly produce the appropriate energy and angle distributions of exit channel particles from 12~MeV proton bombardment, which is  more accurate than relying on internal MCNP physics models for ion-induced channel-based reactions at these relatively low ion bombardment energies.

\begin{figure}
  \centering
  \includegraphics[width=0.48\textwidth]{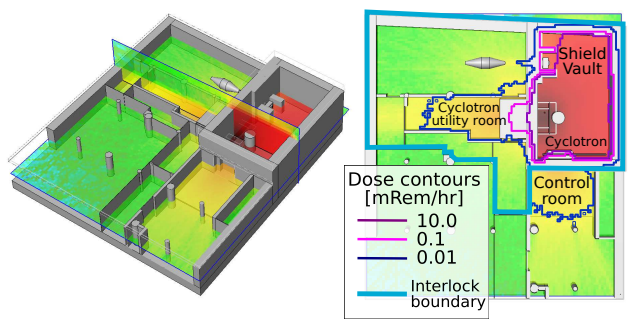}
  \caption{(Left) Three dimensional dose maps superimposed on the full
    MCNP geometry. (Right) Plan view of the Vault Laboratory showing
    the dose map, key radiation dose contours, and the boundaries of
    the interlocked zone, which is restricted during cyclotron
    operation.}
  \label{fig:neutronics}
\end{figure}

To evaluate a conservative "worst case" radiation dose distribution in the laboratory, we considered the potential irradiation of a 100~nm thin film of beryllium with 12~MeV protons at \SI{10}{\micro\ampere} of beam current. While this configuration is for different experimental use of the cyclotron than IEPI (neutron and gamma production for nuclear security), it represents a worst-case scenario from a radiation dose standpoint because of the large neutron-production cross section of fast neutrons, which are the most difficult particles to shield and carry a large dose quality factor of 20. Proton irradiation on beryllium produces large amounts of neutron via the $^9$Be(p,n)$^{9}$B
reaction, which has cross section that peaks around 0.55~barn at approximately 11.0~MeV \cite{byrd1983}. We used the MCNP model to compute the radiation dose within the Vault Facility using dose
equivalent factors . The results are shown in Figure~\ref{fig:neutronics}. While the prompt radiation dose is high
within the vault itself, the dose is ten to one hundred times below the standard 1~mRem/hour threshold within the control room and other non-interlocked accessible areas for occupancy by radiation workers. At this dose level, radiation works could continuously occupy and work safely within the laboratory. Following commissioning of the cyclotron, radiation surveys and active area monitoring have measures dose during IEPI at far lower levels than this worst-case scenario, confirming the safety of performing IEPI in the vault laboratory.

\section{Irradiation capabilities}
\label{sec:irradiation}

This section describes the equipment and processes required to achieve
intermediate energy irradiation of material specimens with a 12~MeV
proton beam.  This includes measurement of the proton beam
spatial distributions required to enable uniform irradiation of samples. Additionally, a pair of beam energy measurements were used to set the material specimen thickness for uniform irradiation and easy post-irradiation testing.

\subsection{The Ionetix ION-12SC proton cyclotron}
\label{subsec:cyclotron}

\begin{figure}
  \centering
  \includegraphics[width=0.48\textwidth]{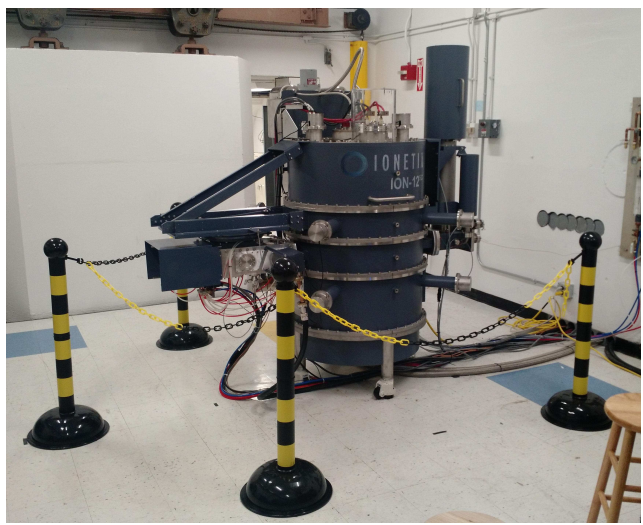}
  \caption{An image of the Ionetix ION-12SC cyclotron installed in the Vault Laboratory}
  \label{fig:cyclotron}
\end{figure}

The core piece of equipment in this facility is the source of intermediate-energy protons: the ION-12SC cyclotron. The ION-12SC, superconducting proton cyclotron, is unique as a high-current, ultra-compact source of protons with energy above 10 MeV. This commercially available  system is nominally capable of producing up to \SI{25}{\micro\ampere} of beam current with a maximum energy of 12.5 MeV.  The superconducting magnets produce a strong magnetic field (4.5 T), allowing a compact machine with an installed weight of only 2.3 tons~\cite{vincent2016}, which can be seen in figure~\ref{fig:cyclotron}.  The specific cyclotron operating at MIT is the first machine of this product line and has been in operation since 2016.

The cyclotron requires a specific set of infrastructure for operation.  As part of the installation, the cyclotron comes with a cryocompressor to cool the superconducting magnets to below 5~K, a water-cooled chiller to remove heat from the compressor and systems internal to the cyclotron (e.g. the ion source and the RF accelerating system), and an electronics rack housing power supplies and monitoring/control equipment. The system needs to be powered (10s of KW), supplied with chilled water, and requires radiation shielding around the cyclotron to avoid risk to personnel. Each of these capabilities was present in the Vault laboratory prior to the arrival of the cyclotron and required minimal restructuring of the space to accommodate this equipment.

There are several features that make this cyclotron well suited for an irradiation facility. As a first-of-its kind ultra-compact cyclotron, it is the smallest and lowest-maintenance source of intermediate energy (10 -- 30~MeV) protons currently available, enabling rapid (>0.1 DPA/day) bulk irradiation (>\SI{100}{\micro\meter}) of materials\cite{jepeal2020}. As an isochronous cyclotron, the proton beam pulses at radio-frequencies (RF) (nominally 68 MHz), acting like a continuous beam for experimental purposes. High achievable beam currents (\SI{10}{\micro\ampere} or more regularly achieved) allows accelerated damage rates during materials irradiation. Finally, the presence of the magnetic field efficiently eliminates surface contamination by carbon ions, a substantial confounding factor in many irradiation studies using ion beams \cite{Shao2019}.

\subsection{Beam spatial distribution}
\label{subsec:imaging}
A strong understanding of the location and shape of the proton beam is essential to ensuring uniform irradiation of sample materials. To this end, the cyclotron beam was imaged over a series of experiments, in order enable the design and execution of subsequent irradiation experiments. 

The core component of the beam imaging experiment is a beam target that scintillates during interaction with the beam. This target is made of a thin sheet of fused quartz (e.g. a microscope slide), mounted on a movable arm that was inserted into the cyclotron. The quartz was mounted at a 45 degree angle to the ion beam, and was monitored by a ccd camera external to the cyclotron. 

\begin{figure}
  \centering
  \includegraphics[width=0.48\textwidth, trim=50 0 0 0, clip]{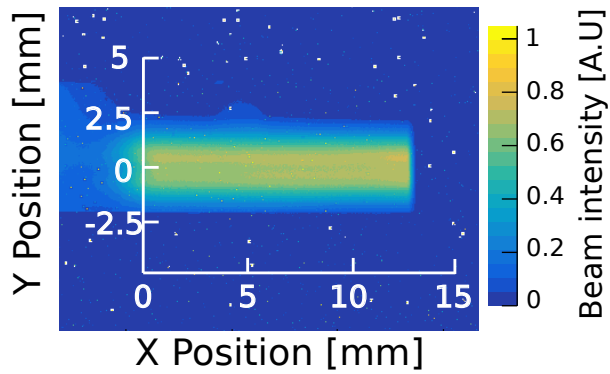}
  \caption{Imaging of the optimized cyclotron beam spot for irradiation on a quartz scintillation target. The color scale in arbitrary Unitas represents the total amount of scintillation light reaching the camera from the quartz target and is, provided no saturation of the image, an excellent measure of beam current distribution and intensity. The 3~mm $\times$ 1~mm gauge region of the tensile specimen shown in Figure~\ref{fig:specimen} is positioned in the center of the yellow uniform beam area.}
  \label{fig:imaging}
\end{figure}

These beam imaging experiments yielded insight into the shape and location of the beam, as well as the degree to which the beam shape could be controlled. In figure \ref{fig:imaging}, an image of the beam is shown with a desired beam shape. From this image, we can see a beam shape that is approximately uniform in a 12mm $\times$ 2mm region. As a result of this, a beam irradiation target was designed to ensure that the tensile gauge region ($\sim$ 3~mm $\times$ 1~mm) to fit easily within the uniform region. This beam shape is dictated by the static shaping of the cyclotron's internal magnetic field, and therefore  cannot be controlled in-situ and does not change over time ("drift").

\subsection{Beam energy measurement}
Understanding the incident proton energy of the cyclotron beam is critical to the design of effective experiments, with implications in the range of the protons in sample materials, the amount and type of induced radiation, and the characteristics of the radiation damage \cite{jepeal2020}.  While the beam energy is determined by the strength and shape of the magnetic fields, and is therefore fixed for each machine, there is a need to independently confirm the manufacturer-specified beam energy while accounting for deviations that could be specific to this machine.  To meet this need for an accurate representation of the proton energy, two techniques were developed and implemented in the cyclotron under the desired experimental settings. 

One energy measurement technique utilized gamma spectroscopy of long-lived radio-isotopes in copper foils following a well-characterized irradiation. The principle behind this technique is that through determining the amount of a particular nuclear reaction induced at several locations within a sample, the energy of the beam can be inferred from known cross section and stopping power data. This type of technique has been used previously in cyclotron beam energy measurements \cite{Burrage2009, Kopecky1985, Kim2006}, and has been adapted and optimized for this beam energy.

\begin{figure}
  \centering
  \includegraphics[width=0.48\textwidth]{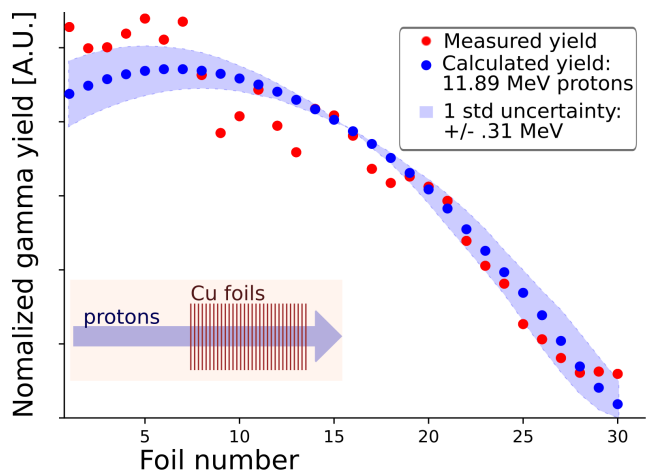}
  \caption{A comparison of the simulated and measured activation of copper foils at the minimum chi-squared value, which results in an incident proton beam energy measurement of 11.89$\pm$0.31 MeV }
  \label{fig:foils}
\end{figure}

In this implementation, the $^{65}$Cu(p,n)$^{65}$Zn reaction was utilized for its principal, easily measured gamma energy \\(1.115~MeV) and long half life product (244 days). A target was constructed of 29 foils of pure copper, each \SI{10}{\micro\meter} thick. These foils were exposed to a proton beam perpendicular to their surface, producing $^{65}$Zn in each foil corresponding to the proton energy in that foil. Several days after the exposure, the foils were placed between two sodium-iodide gamma detectors (4 inch cubes of NaI coupled to 3 inch photomultiplier tubes) allowing consistent, nearly full (4$\pi$) solid angle coverage. The ADAQ Framework \cite{hartwig2016} was used to acquire and analyze gamma spectra from each foil. After background and continuum subtraction, the 1.115~MeV gamma peak was fit with a Gaussian function and integrated to calculate the relative total amount of activity of the $^{65}$Zn in each foil. The 1.115~MeV gamma yield data was compared against calculations of the expected yield using known cross sections and stopping powers. These calculations were performed for a wide range of proton energies, and a chi-squared minimization and error estimation was performed (according to the formula $[0.5\dif^2 \chi^2/\dif E^2]^{-1/2}$), yielding an incident proton energy measurement of \\11.9$\pm$0.3~MeV as shown graphically in Figure~\ref{fig:foils}.

A second beam energy estimation was performed by measuring the yields of the gamma lines from (p,p$^\prime\gamma$) nuclear reactions. These (p,p$^\prime$) reactions have large negative $Q$-values, resulting in an energetic threshold which is within a few MeV of the 12 MeV proton energy.  These thresholds makes the yield for a particular gamma line strongly dependent on the incident beam energy. Known (p,p$^\prime$) cross sections were used with simulations of detector response to produce simulated spectra at a range of beam energies. These simulated spectra were compared to the measured spectrum, and the best fit simulation yielded the estimated beam energy. To the best of our knowledge, this is the first time such an in-situ nuclear reaction-based technique has been used to infer the energy of a cyclotron beam.

\begin{figure}
  \centering
  \includegraphics[width=0.48\textwidth,clip]{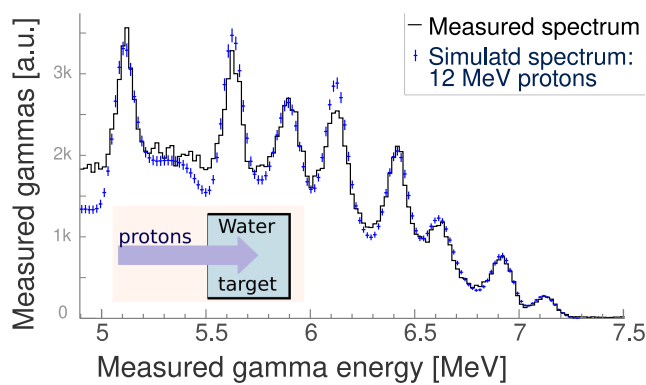}
  \caption{Comparison of the simulated and experimental gamma spectra at the minimum of the chi-squared fit comparison, which results in an incident proton beam energy of 12$\pm$0.3~MeV.}
  \label{fig:spectrum}
\end{figure}

In this experiment 6.13, 6.92, and 7.12 MeV gamma rays were generated by protons through  \\$^{16}$O(p,p$^{\prime}\gamma$)$^{16}$O reactions in a water target. These gammas were measured with a LaBr$_3$(Ce) detector, using the ADAQ framework~\cite{hartwig2016} to acquire a gamma spectrum.  The detector response to each of these gamma energies was then simulated using the Geant4 toolkit~\cite{geant4}, accounting for the geometry, the detector efficiencies, and the detector resolution.  The relative yields of each gamma energy were calculated for proton energies of 8-14 MeV with 0.1 MeV increments with known cross sections~\cite{1981DY03,1998_all_O} and stopping powers~\cite{pstar}, using the following equation: $Y = -\frac{\rho N_{\text{Av}}}{A} \int_{E_{\text{th}}}^{E_{\text{p}}} \dif E [\sigma(E)/{\frac{\dif E}{\dif x}}] $, where $\rho$ is the target density, $N_{\text{Av}}$ is Avogadro's number, $A$ is the atomic mass, ${\frac{\dif E}{\dif x}}$ is the stopping power, $\sigma(E)$ is the production cross section, $E_{\text{th}}$ is the production threshold, and $E_{\text{p}}$ is the beam energy. By adding the simulated detector responses for each gamma energy with weights determined by the calculated yields, a simulated gamma spectrum was produced for each assumed beam energy value. Each simulated spectrum was normalized by total number of counts and compared to the experimental spectrum in a chi-squared test.  The uncertainty of the inferred beam energy was determined as above, producing a proton energy estimate of 12.0$\pm$0.3 MeV.  Figure~\ref{fig:spectrum} presents the simulated spectrum with the minimum chi-squared value.  The fit-data deviations at $\sim$5 MeV are due to the proton bremsstrahlung~\cite{nattress2019characterization} not included in the simulation model. 

The agreement between two independent experimental measurements (11.9$\pm0.3$MeV and 12.0$\pm0.3$MeV) of the beam energy is provides confidence in designing beam targets and analyzing beam interactions in matter during irradiation.  One primary use of this exact beam energy measurement is in the the selection of the maximum sample thickness that enables uniform irradiation without proton implantation in the sample. This exact thickness is dependent on specific material properties, but, in the case of 12 MeV protons, has been demonstrated to vary 150-300 um for common metals used in nuclear materials\cite{jepeal2020}. 

\subsection{Irradiation target assemblies}
\label{subsec:target}

In order to enable effective irradiation experiments, a number of capabilities are needed, including: 1) alignment of a sample within the uniform beam region, 2) monitoring of the amount of exposure to the ion beam, 3) control of temperature during experiments, and 4) uniform temperature throughout the sample during the experiment. In order to meet these criteria, a custom irradiation target was designed and implemented.

\begin{figure}
  \centering
  \includegraphics[width=0.5\textwidth]{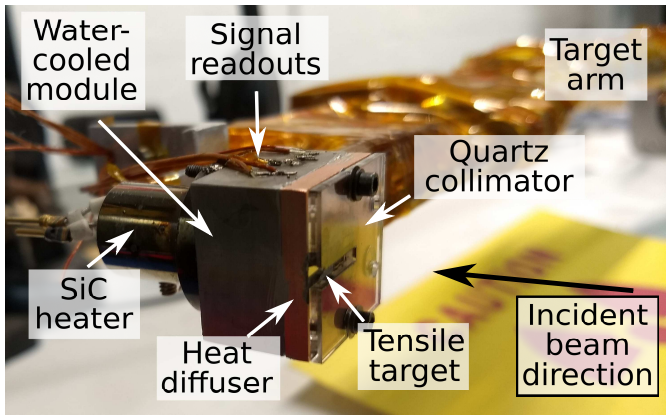}
  \caption{The internal cyclotron beam target assembly used for
    materials irradiation.}
  \label{fig:target}
\end{figure}

Alignment of the target within the proton beam is critical to a uniform irradiation and a valid measurement of the material properties. In this irradiation target, alignment is achieved through two features: imaging of the beam during experiments and precision adjustment of the target location.  The beam is continuously imaged during experiments using a custom quartz collimator (shown in figure \ref{fig:target}), on which the proton beam is viewed by a camera as described in section \ref{subsec:imaging}. The target is inclined at 10 degrees relative to the proton beam in order to enable line-of-sight imaging to the camera external to the cyclotron. The precision adjustment is achieved through a sensitive tilting mechanism (an "alignment gimbal" purchased from Nor-Cal products), capable of positioning the target with a precision within \SI{100}{\micro\meter}. During operation, the target is adjusted while imaging a very low beam current (<\SI{100}{\nano\ampere}), until the beam surrounds the aperture in the collimator, ensuring a uniform exposure to the sample within the aperture. After the target is aligned, the beam can be brought up to its full intensity (\SI{1}{\micro\ampere}-\SI{10}{\micro\ampere}).

Monitoring of the exposure to the ion beam requires the target to act as a Faraday cup, allowing measurement of the electric current produced by the ion beam.  To achieve this, the target is electrically isolated from the rest of the equipment with the exception of a single wire that delivers the current to a nano-ammeter for measurement. The presence of a strong magnetic field ($\sim$4~T) acts as a built in secondary electron suppression, eliminating the need biasing the irradiation target \cite{Podadera2013}.

Controlling the sample temperature during irradiation requires balancing the heat input by the ion beam against the heat removal through conduction.  For example, using a typical beam currents of \SI{10}{\micro\ampere} requires managing greater than 100 W of heat deposition in a relatively small volume (12~mm $\times$ 2~mm $\times$ 0.3~mm). In order to offset this heat input and allow a steady-state operating temperature, a water-cooled module is included in the target design.  Through control of the ion beam intensity, this system is capable of maintaining a fixed temperature within $\sim$5 K. In case an experiment requires decoupling the sample temperature from the ion beam intensity, a ceramic heater is also included in the target design, allowing an external source of heat to balance against the cooling module. 

\begin{figure}
  \centering
  \includegraphics[width=0.5\textwidth]{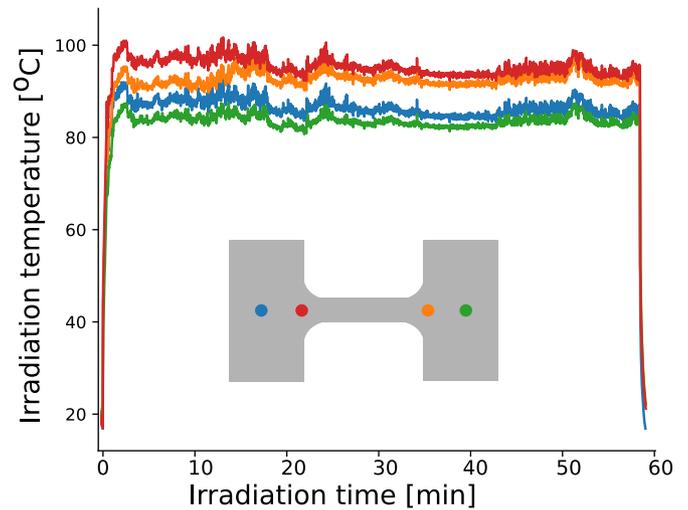}
  \caption{Sample irradiation temperature uniformity measured by welded thermocouples.}
  \label{fig:IrrTemp}
\end{figure}

Enabling uniform temperature within a sample is a challenge that requires uniform heat deposition by the ion beam, uniform thermal contact between the sample and its holder, and precise monitoring of the temperature at multiple locations within the sample.  Uniform thermal contact is achieved by the use of 1) uniform pressure on the sample, achieved by screw-tightened springs at each of the four corners of the sample and 2) a thin, compressible material between the sample and the holder, such as a deformable foil (e.g. silver).  Monitoring of the local temperature variations is achieved directly by the mounting of small (\SI{250}{\micro\meter} diameter) thermocouples within the exposed area on either end of the gauge region of the sample.  Four thermocouple are typically used ensure that there are no substantial temperature gradients and to allow redundancy in case of thermocouple failure. A typical temperature curve can be seen in figure \ref{fig:IrrTemp}.

\subsection{Irradiation tensile specimens}
\label{subsec:specimen}

Successful mechanical testing of irradiated samples requires a custom tensile specimen geometry. This geometry must allow for uniform irradiation within the cyclotron beam and bulk measurement of tensile properties after irradiation.

\begin{figure}
  \centering
  \includegraphics[width=0.5\textwidth]{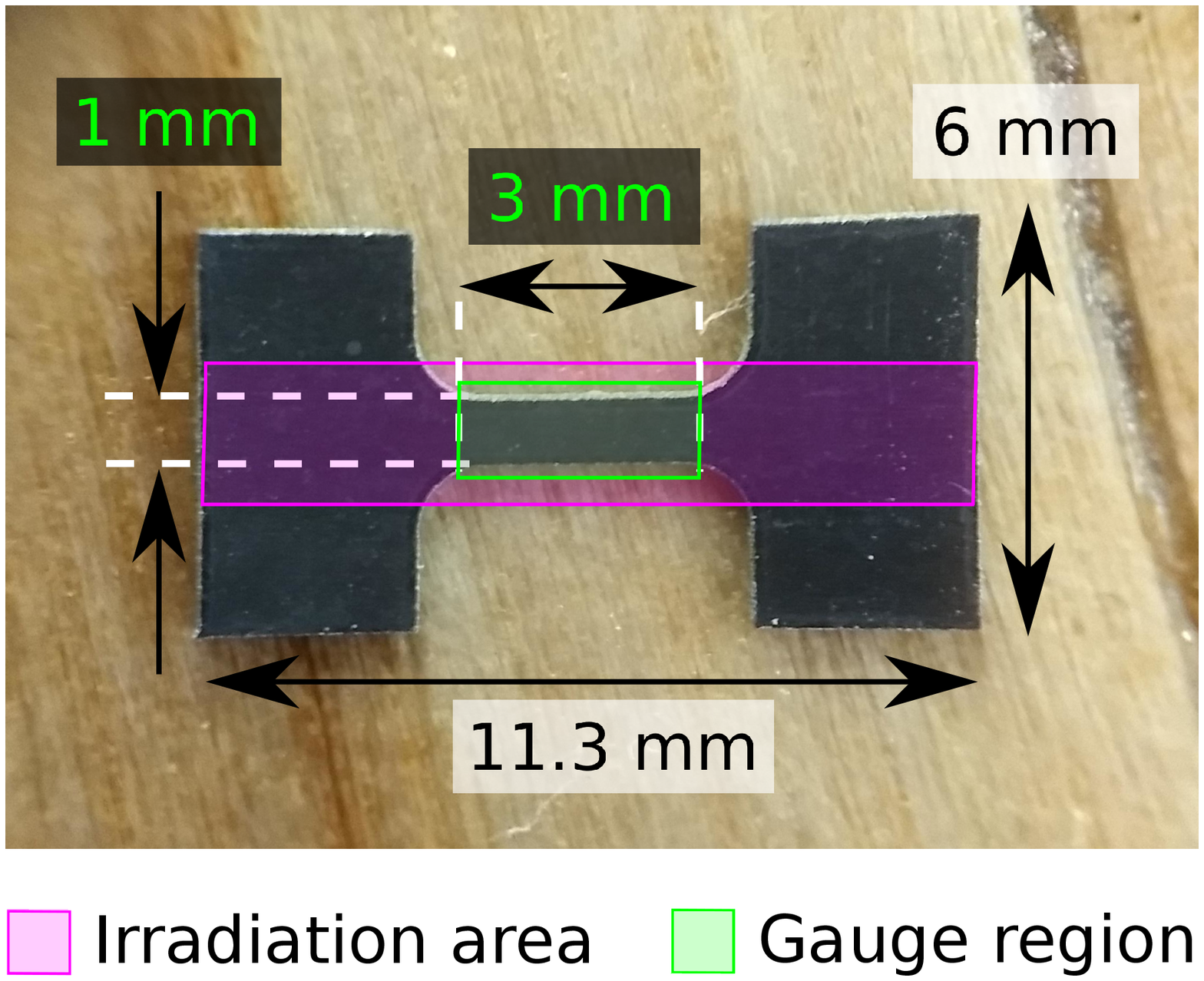}
  \caption{The custom tensile specimen that is mounted to the target
    assembly for irradiation within the cyclotron. The cyclotron beam
    is collimated to uniformly irradiate the (purple) irradiation area, including the (green) gauge region.}
  \label{fig:specimen}
\end{figure}

Uniform irradiation of tensile specimens requires both that the 2D profile of the tested region fit within the uniform beam size of the cyclotron and that the thickness not exceed the uniform damage range for the protons. As shown in figure \ref{fig:specimen}, a 3~mm $\times$ 1~mm gauge region was designed to allow easy alignment within the 12~mm $\times$ 2~mm beam spot. 

The thickness of tensile samples is allowed to vary by application within the range of 200 -- \SI{300}{\micro\meter}, enabling uniform irradiation without implantation of protons, as described in a separate publication~\cite{jepeal2020}.  Previous work has demonstrated that miniature samples with thickness as small as \SI{200}{\micro\meter} are representative of larger samples, with nearly equal values of yield strength, ultimate tensile strength, and uniform elongation.\cite{Gussev2014,Takeda2017}.  Size effects are present in radiation damage accumulation, but these affects are limited to much smaller volumes, typically less than \SI{1}{\micro\meter}\cite{zinkle2018}.  Therefore, sample thickness in the range of 200 -- \SI{300}{\micro\meter} can enable both, uniform radiation damage and accurate tensile testing that is representative of bulk materials.

The grip and shoulder regions of the sample geometry, shown in figure \ref{fig:specimen} are designed to 1) concentrate deformation within the gauge region and 2) allow a greater surface area for transferring force from the tensile tester to the sample. The smaller width of the gauge region and the smooth curvature of the shoulders create a high, uniform stress area in this gauge region. As shown in figure \ref{fig:strain}, this design allows uniform deformation within the gauge region, until the end of the test, when strain localization occurs and leads to mechanical failure.  Irradiation hardening can occur both in the gauge region and in the grip and shoulder regions, but does so uniformly, maintaining a preference to deform and eventually fracture in the gauge region.  

\section{Post-irradiation experimental equipment}
\label{sec:pie}

This section highlights the key equipment presently available within the Vault Laboratory for materials testing following the irradiation within the cyclotron. Nearby facilities at MIT that are not described here provide additional post-irradiation capabilities, such as nano- and micro-indentation as well as SEM/TEM imaging. An advantage of the IEPI technique is that the low residual activation of the irradiated samples, especially after a few week dwell time, can enable access to powerful equipment and facilities that is typically unable or unwilling to handle highly activated samples such as those produced in neutron irradiations.

\subsection{Tensile tester}
\label{subsec:tester}

The tensile tester system was designed and constructed as a means to measure changes to bulk mechanical properties in IEPI samples.  In order to do so, the system must apply uni-axial tensile force to deform miniature samples, concurrently measure the strain of these samples,  and (for many applications) elevate the sample temperature and control its environment during testing. 

\begin{figure}
  \centering
  \includegraphics[width=0.48\textwidth]{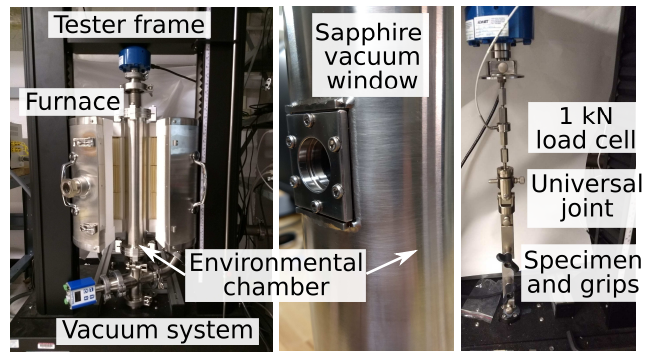}
  \caption{The high temperature ($\lesssim$1000$^{o}$~C), high vacuum
    ($\lesssim$10$^{-6}$~torr) tensile test assembly. The retort is
    surrounded by a furnace and provides a vacuum environment and
    thermal transfer to the internal tensile specimen. Concentric
    windows in the furnace and retort provide visual access to the stain measurement
    system.}
  \label{fig:tensile}
\end{figure}

The force and displacement for mechanical testing is generated by a tensile test frame: the eXpert2611 test frame prodcued by ADMET Inc.  This system is capable of producing 10~kN of force, traveling over 1.2~m at speeds up to 1~m/min.  This model was chosen primarily because of its open, two-column design, leaving room for additional equipment as shown in figure~\ref{fig:tensile}. Attached to the test frame are a series of small components, including load cells (for measuring force), grips (for transferring force to the sample), and universal joints (for alignment).  While this equipment can span a range of forces, a typical test of the tensile specimen shown in figure \ref{fig:specimen} requires only 100~N -- 1~kN, and utilizes load cells and grips appropriate for this. 

\subsection{Non-contact strain measurement}
\label{subsec:DIC}

\begin{figure}
  \centering
  \includegraphics[width=0.48\textwidth]{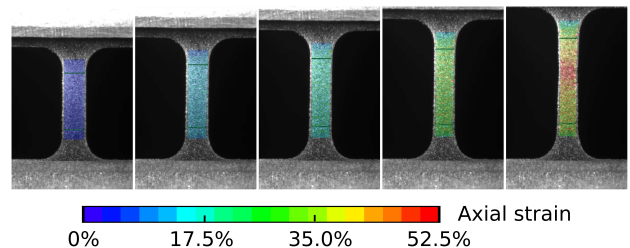}
  \caption{A tensile test of a custom specimen performed by the
    digital image correlation (DIC) system, viewing the sample through
    the furnace and retort window. A computed strain color map is
    automatically superimposed; two horizontal black lines on each
    sample shown as a virtual extensometer.}
  \label{fig:strain}
\end{figure}

Because of the miniature sample size and the need to test at high temperatures, a non-contact, image-based strain measurement system was implemented.  In this system, a usb camera (mvBlueFOX3) is used to take images of a sample during testing, as demonstrated in figure \ref{fig:strain}.  These images are converted into strain measurements through the free "Correlate" software produced by the company GOM in a process known as digital image correlation.  This system is a direct replacement for a traditional, clip-on strain measuring device, and has been validated to produce data consistent with such devices.

\subsection{Environmental control}
\label{subsec:furnace}

For high temperature measurements, a heating and environmental control system has been developed.  A Thermcraft Incorporated split-tube furnace is mounted on the tensile tester and can be closed around a tensile sample during testing. This furnace is capable of reaching high temperatures, up to 1200~$^o$C, and includes a view-port to allow optical strain measurements to be made from outside of the furnace. An environmental chamber fits within the furnace, allowing a tested sample to be heated under vacuum ($10^{-6}$ torr) or an inert atmosphere to prevent corrosion. This chamber includes a custom made, high temperature view port capable of maintaining vacuum and allowing strain measurements. 

\subsection{Validation of mechanical tests}
\label{subsec:validation}

A series of tensile tests were performed to qualify the customized miniature tensile specimen and validate the ability of the tensile testing system to produce accurate and reproducible strength and ductility data. \SI{300}{\micro\meter} thick sheets of solution-annealed Alloy 625 were acquired from Goodfellow Inc. These sheets were electrical discharge machined into tensile specimen with the geometry shown in figure \ref{fig:specimen}. These specimens were tested in air at room temperature with the tensile system depicted in figure \ref{fig:tensile} at a constant strain rate. The DIC system was used to create a virtual strain gauge, measuring average strain across the sample. Five such tests were performed from the same sheet of stock material and compared to ensure the tensile measurements were accurate and reproducible.

\begin{figure}
  \centering
  \includegraphics[width=0.48\textwidth]{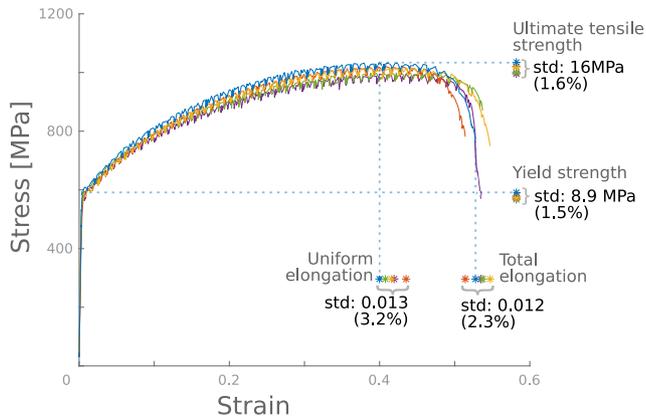}
  \caption{A validation of the tensile testing system using optical strain measurement (DIC) demonstrating high reproducibility for strength and ductility measurements}
  \label{fig:ttest}
\end{figure}

As shown in figure \ref{fig:ttest}, the tensile testing system demonstrated a high degree of reproducibility and accuracy.  In both ultimate tensile strength and yield strength, the measurements were reproducible to within 2\% of the absolute value. Similarly the uniform elongation and total elongation were reproducible to within 0.02 strain. The mean ultimate tensile strength was measured to be 1010 MPa, which agrees with the manufacturer specification of 950 MPa.  Additionally, the DIC system presented uniform strain along the gauge region of the tensile samples as depicted in figure \ref{fig:strain}. These results qualify the customized miniature tensile specimen and ensure that the specimen, combined with the tensile tester and DIC system, provide accurate, precise, and repeatable tensile measurements. It ensures that changes in tensile properties between pristine (unirradiated) and irradiated materials are due to radiation-induced evolution of the material microstructure.

\section{Conclusion}
\label{sec:conclusion}

We have provided an overview of the first accelerator facility dedicated to performing rapid, bulk materials irradiation with a \SI{10}{\micro\ampere}, 12~MeV proton beam and post-irradiation tensile testing to extract engineering material properties. The irradiation is enabled by an Ionetix ION-12SC 12~MeV compact superconducting proton cyclotron with customized internal target assemblies to ensure uniform beam on the sample and minimal temperature gradients. The tensile testing is performed using a custom, mm-scale sample geometry with thicknesses of 100-\SI{300}{\micro\meter}. A specialized environmental chamber can be used in the tensile test to provide high temperature (up to 1000~$^{o}$C) and/or high vacuum (<10$^{-6}$~torr) conditions (or other custom environments) when needed. A digital image correlation system provides non-contact strain measurements through high temperature windows in the furnace and environmental chamber. The first results from the IEPI technique for nuclear materials (published elsewhere) are compelling, indicating that rapid irradiation and materials testing with high fidelity to advanced fission and fusion energy systems is achievable \cite{jepeal2020}. Therefore, IEPI promises to be an effectively materials evaluation technique that can be deployed widely to maximize progress in researching, down-selecting, and qualifying materials for advanced nuclear application.

The key enabling technology for IEPI is commercially available, compact, superconducting cyclotrons, which expand access to 10 -- 30~MeV protons beyond the fields of high energy physics and isotope generation.  Compared to research reactors, IEPI facilities for nuclear materials research can now be established with far lower capital and operating costs, significantly smaller laboratory footprint, and less safety and activated sample handling equipment. The increasing use of such machines could mirror the nearly ubiquitous use of low-energy ion accelerators in university labs (e.g. tandems and Van de Graaffs) that have accelerated research in a variety of fields and created new commercial applications for ion beams.  We expect the growth of IEPI to complement irradiation in nuclear reactors and low-energy ion beams and to accelerate the development of materials for future of fission and fusion energy.

\section{Acknowledgements}
The authors are indebted to the following people for their contributions to setting up this facility and making the research work possible: Lance Snead for his technical insights on materials irradiation science; Alberto Pontarollo, Maria Elena Gennaro, and their colleagues at Eni for their guidance and management of the research; Kevin Woller and Pete Stahle for their assistance in keeping the Vault's accelerators running smoothly; Tim Schmidt from Trilion Quality Systems for his guidance in developing the DIC system; Ken Stevens for leading the installation, commissioning, and initial operations of the cyclotron; Jay Paquette, Dan Alt, Gary Horner, and the team at Ionetix for the their technical support of the cyclotron; and Guy Muzio, Barry Turkanis, Marcia Sajewicz, and John Haines for enabling the cyclotron to be at MIT; 

Several of the authors have moved to new institutions since this work was completed: Enrique Velez Lopez is now at MIGHTR LLC.; Nick Schwartz is now at the University of Illinois Urbana-Champaign; Brandon Sorbom is now at Commonwealth Fusion Systems

This work was funded by (1) The MIT Research Support Committee and (2) Eni S.p.A. through the MIT Energy Initiative and the Laboratory for Innovation in Fusion Technology (LIFT) at the MIT Plasma Science and Fusion Center.

\printcredits


\bibliography{references}



\end{document}